\documentclass[aps,prd,preprint,amsmath,showpacs]{revtex4-1}
\usepackage{graphicx}
\usepackage{dcolumn}
\usepackage{bm}
\usepackage{verbatim}

\newcommand{\Det}{{\rm Det}}

\newcommand{\UU}{{\cal U}}
\newcommand{\DD}{{\cal D}}
\newcommand{\NN}{{\cal N}}
\newcommand{\LL}{{\cal L}}
\newcommand{\No}{{\cal T}_{\scriptscriptstyle +0}}
\newcommand{\Lo}{{\cal T}_{\scriptscriptstyle 0+}}
\newcommand{\Ni}{{\cal T}_{\scriptscriptstyle -0}}
\newcommand{\Li}{{\cal T}_{\scriptscriptstyle 0-}}
\newcommand{\Nd}{{\cal D}_{\cal U}}
\newcommand{\Ld}{{\cal D}_{\cal D}}
\newcommand{\NL}{{\cal T}_{\scriptscriptstyle ++}}
\newcommand{\NiLi}{{\cal T}_{\scriptscriptstyle --}}
\newcommand{\NLi}{{\cal T}_{\scriptscriptstyle +-}}
\newcommand{\NiL}{{\cal T}_{\scriptscriptstyle -+}}
\newcommand{\GU}{G_U}
\newcommand{\GD}{G_D}
\newcommand{\Iqi}{{\cal I}^q_{\scriptscriptstyle TD}}
\newcommand{\Iqj}{{\cal I}^q_{\scriptscriptstyle PD}}
\newcommand{\Iqptd}{(\Iqj/\Iqi)}
\newcommand{\Iqk}{{\cal I}^{ql}_{\rm prod}}
\newcommand{\Ili}{{\cal I}^{\ell}_{\scriptscriptstyle TD}}
\newcommand{\Ilj}{{\cal I}^{\ell}_{\scriptscriptstyle PD}}
\newcommand{\Iqv}{{\cal I}^q_{\scriptscriptstyle DV}}
\newcommand{\Ia}{{\cal I}^{ql}_{\rm comm}}
\newcommand{\Ib}{{\cal I}^g_{12}}
\newcommand{\Ic}{{\cal I}^g_{13}}
\newcommand{\mat}[9] {\left ( \begin{matrix} #1 #2 #3 \\
                                     #4 #5 #6   \\
                                     #7 #8 #9       \end{matrix} \right ) }
\newcommand{\colvec}[3] {\left( \begin{matrix} #1  \\
                                        #2  \\
                                        #3  \end{matrix} \right)}
\newcommand{\rowvec}[3] {\left( \begin{matrix} #1, & #2, & #3 \end{matrix} \right)}

\begin{document}

\preprint{RAL-TR-2010-020}

\title{Exact One-Loop Evolution Invariants in the Standard Model}

\author{P.~F.~Harrison}
\email[]{p.f.harrison@warwick.ac.uk}
\author{R.~Krishnan}
\email[]{k.rama@warwick.ac.uk}
\affiliation{Department of Physics, University of Warwick, Coventry CV4 
7AL, UK}
\author{W.~G.~Scott}
\email[]{w.g.scott@rl.ac.uk}
\affiliation{Rutherford Appleton Laboratory, Chilton, Didcot, 
Oxon OX11 0QX, UK}

\date{\today}

\begin{abstract}
Guided by considerations of flavour symmetry, we construct a set of exact Standard Model (SM) renormalisation group evolution invariants which link quark masses and 
mixing parameters. We examine their phenomenological implications and infer a simple 
combination of Yukawa coupling matrices which plays a unique role in the 
SM, suggesting a possible new insight into the observed spectrum of quark masses. Our evolution invariants are readily generalised to the leptons in the case of Dirac
neutrinos, but do not appear to be relevant for either quarks or leptons in the MSSM.
\end{abstract}

\maketitle

\section{Introduction}
Recently, there has been interest in evolution invariants \cite{Demir,Liu,Xing,Kuo}, combinations of observables which do not evolve under the renormalisation group (RG). Usually, they have been derived in various approximations, eg.~assuming no fermion mixing \cite{Demir}, or assuming only two generations of fermions \cite{Liu}, or neglecting the  contributions of light quark masses \cite{Xing}. Many work beyond the SM \cite{Demir,Kuo}.
The RG evolution equations (RGEs) of the Yukawa couplings are compactly written as matrix equations, since the problem is intrinsically flavour-symmetric - all flavours are treated equivalently. Conventional flavour observables, such as  the quark and lepton masses (proportional to the eigenvalues of the Yukawa coupling matrices) or their mixing angles, break the flavour symmetry so that their RG equations are more complicated \cite{Babu,Xing}. Motivated by our earlier work on flavour-symmetric variables \cite{flavinv,*FCGA}, we construct a set of flavour-symmetric observables whose one-loop RG equations in the SM are rather simple. These lead straightforwardly without approximation to new evolution invariants which are exact at this order. For illustration, we consider primarily the quarks, but our considerations are equally valid for the leptons in the case that neutrinos are Dirac particles, in which case more invariants follow.

We define the Hermitian squares of the Yukawa coupling matrices for charge $+\frac{2}{3}$ ($U$) and charge $-\frac{1}{3}$ ($D$) quarks respectively:
\begin{align}\label{eq:hermsq}
\UU & = U^\dagger U, & \DD & = D^\dagger D
\end{align}
and introduce a complete set of ten flavour-symmetric invariants (each is invariant under independent $S3$ permutations of the $(u,c,t)$ and/or the $(d,s,b)$ flavour labels):
\begin{align}\label{eq:inv}
\No &= Tr(\UU) & \Lo &= Tr(\DD) \notag \\
\Ni &= Tr(\UU^{-1}) & \Li &= Tr(\DD^{-1}) \notag \\
\NL &= Tr(\UU\DD) & \NLi &= Tr(\UU\DD^{-1})  \\
\NiL &= Tr(\UU^{-1}\DD) & \NiLi &= Tr(\UU^{-1}\DD^{-1}) \notag \\
\Nd &= \Det(\UU) & \Ld &= \Det(\DD).\notag
\end{align}
The set is complete in the sense that the ten variables are  fully determined by the physical masses and mixings, and are in turn, sufficient to fully determine them (up to discrete permutations of the flavour labels). A further ten  analogous variables can be similarly constructed using Hermitian squares of Yukawa matrices for the neutrinos ($\NN$) and the charged leptons ($\LL$).

\section{SM Evolution}
We start with the one-loop RG equations for the quark Yukawa coupling matrices in the SM \cite{Machacek2}: 
\begin{align}
\label{eq:rg1}
U^{-1} \frac{dU}{dt} &= \gamma_u + \frac{3}{2}(\UU-\DD),\\
\label{eq:rg2}
D^{-1} \frac{dD}{dt} &= \gamma_d + \frac{3}{2}(\DD-\UU)
\end{align}
where $t=\frac{1}{16\pi^2}\ln{(\mu/\mu_0)}$ for renormalisation-scale $\mu$, and
\begin{equation}
\gamma_u=T-\GU;\qquad\gamma_d=T-\GD,\label{Eq:gammas}
\end{equation}
with:
\begin{align}
T  &= Tr(3\,\UU + 3 \DD + \NN + \LL), \label{Eq:T}\\
\GU  &= \frac{17}{12} g_1^2 + \frac{9}{4} g_2^2 + 8 g_3^2, \label{Eq:GU}\\
\GD  &= \frac{5}{12} g_1^2 + \frac{9}{4} g_2^2 + 8 g_3^2\label{Eq:GD}
\end{align}
(the $g_i$ are the gauge couplings \footnote{Throughout, we use the $U(1)$ gauge coupling normalisation of \cite{Babu}.}). For $\UU $ and $\DD$, the Hermitian-squared matrices of Eq.~(\ref{eq:hermsq}), we get:
\begin{align}
\label{eq:hermsqrg1}
\frac{d\UU}{dt}&= 2\gamma_u\UU + 3 \UU^2 - \frac{3}{2} \{\UU, \DD\}, \\ 
\label{eq:hermsqrg2}
\frac{d\DD}{dt}&= 2\gamma_d\DD + 3 \DD^2 - \frac{3}{2} \{\UU, \DD\}.
\end{align}

Differentiating Eqs.~(\ref{eq:inv}) and using Eqs.~(\ref{eq:hermsqrg1}) and (\ref{eq:hermsqrg2}), we obtain the separate evolution equations of our ten flavour-symmetric observables:
\begin{subequations}
\begin{align}
&\frac{d\No}{dt}= 2\gamma_u\No+3(\No^2-2 \Ni \Nd - \NL) \label{eq:ourset1}\\
&\frac{d\Lo}{dt}= 2\gamma_d\Lo+3(\Lo^2-2 \Li \Ld - \NL) \label{eq:ourset2}\\
&\frac{d\Ni}{dt}=-2\gamma_u\Ni-9+3 \NiL\label{eq:ourset3}\\
&\frac{d\Li}{dt}=-2\gamma_d\Li-9+3 \NLi\label{eq:ourset4}\\
&\frac{d\NL}{dt}= 2(\gamma_u+\gamma_d) \NL\label{eq:ourset5}\\
&\frac{d\NiLi}{dt}=-2(\gamma_u+\gamma_d) \NiLi\label{eq:ourset6}\\
\begin{split}
&\frac{d\NLi}{dt}= 2 (\gamma_u-\gamma_d+3\No)\NLi-6\No\\
& \text{ \qquad \quad} +6\Nd(\NiLi-\Ni\Li)\label{eq:ourset7}\\
\end{split}\\
\begin{split}
&\frac{d\NiL}{dt}= 2 (-\gamma_u+\gamma_d+3\Lo)\NiL-6\Lo\\
& \text{ \qquad \quad} +6\Ld(\NiLi-\Ni\Li)\label{eq:ourset8}\\
\end{split}\\
&\frac{d\Nd}{dt}= 3 \Nd[2\gamma_u+(\No-\Lo)]\label{eq:ourset9}\\
&\frac{d\Ld}{dt}= 3 \Ld[2\gamma_d-(\No-\Lo)]\label{eq:ourset10}.
\end{align}
\end{subequations}

We make the following observations:
\begin{itemize}
\item Most of the variables' evolutions have two parts: 
\begin{itemize}
\item a part proportional to the variable itself, whose coefficient depends at most on $\gamma_u$, $\gamma_d$, $\No$ and $\Lo$. We call this the ``pure'' part. 
\item a part which depends more generally on the other variables - the ``mixed'' part.
\end{itemize}
\item The four variables $\Nd$, $\Ld$, $\NL$ and $\NiLi$ have only pure parts (this is also the case for Jarlskog's determinant \cite{Jarlskog1,*Jarlskog2}, which was the main result of Ref.~\cite{Athanasiu}). This feature seems to be peculiar to the SM - we will rely on it in the next stage of our derivation.
\end{itemize}

\section{SM Evolution Invariants}
Exploiting the opportunity to cancel the terms involving $\No$ and $\Lo$ in Eqs.~(\ref{eq:ourset9}) and (\ref{eq:ourset10}), we note that the quantity $\Det(\UU\DD)=(\Nd\Ld)$ has a pure evolution with exactly a factor three times the coefficient which appears in Eqs.~(\ref{eq:ourset5}) and (\ref{eq:ourset6}):
\begin{equation}\label{eq:UDevol}
\frac{d}{dt}\ln{\Det(\UU\DD)}=6(\gamma_u+\gamma_d).
\end{equation}
We may thus form two independent combinations which are exact evolution invariants at one loop order \footnote{Our notation for $\Iqi$ and $\Iqj$ is based on which of the coefficients of the eigenvalue equation of the matrix $\UU\DD$ the RG invariant may be constructed from, eg. $\Iqj=P(\UU\DD)/\Det^{\frac{2}{3}}(\UU\DD)$, where $P(\UU\DD)$ is defined in Eq.~(\ref{Eq:Pdef}).
}:
\begin{gather}
\Iqi\equiv\frac{\NL}{(\Nd\Ld)^{\frac{1}{3}}}\equiv\frac{Tr(\UU\DD)}{\Det^{\frac{1}{3}}(\UU\DD)};\qquad\frac{d\Iqi}{dt}=0\label{eq:Iq1},\\
\Iqj\equiv\NiLi (\Nd\Ld)^{\frac{1}{3}}\qquad\qquad\qquad\qquad\qquad\,\,\,\,\,\nonumber\\
\equiv Tr(\UU\DD)^{-1}\Det^{\frac{1}{3}}(\UU\DD);\qquad\frac{d\Iqj}{dt}=0\label{eq:Iq2}.
\end{gather}

The pure evolutions expressed by Eqs.~(\ref{eq:ourset5}), (\ref{eq:ourset6}) and (\ref{eq:UDevol}), and the two resulting RG invariants, Eqs.~(\ref{eq:Iq1})-(\ref{eq:Iq2}), are the key results of this paper. $\Iqi$ and $\Iqj$ appear to be the only exact RG invariants that can be constructed from the quark Yukawa coupling matrices alone in the SM case. We have not succeeded in finding  similar exact RG invariants involving only Yukawa couplings in the MSSM or the 2 Higgs doublet model (2HDM).

We can construct entirely analogous evolution invariants using $\NN$ and $\LL$, the (Hermitian squares of the) Yukawa coupling matrices for the leptons (in the Dirac neutrino case). The RGEs of $\NN$ and $\LL$ are analogous to those in Eqs.~(\ref{eq:hermsqrg1})-(\ref{eq:hermsqrg2}) with $\gamma_{\nu}$ and $\gamma_{\ell}$ defined as in  Eq.~(\ref{Eq:gammas}) with the same value of $T$ (Eq.~(\ref{Eq:T})) and the gauge contributions,  Eqs.~(\ref{Eq:GU}) and (\ref{Eq:GD}), modifed to $G_N=\frac{3}{4}g_1^2 + \frac{9}{4} g_2^2$ and $G_L= \frac{15}{4}g_1^2 + \frac{9}{4}g_2 ^2$. The leptonic analogue of the ``pure'' evolution rate $2(\gamma_u+\gamma_d)$, Eqs.~(\ref{eq:ourset5}-\ref{eq:ourset6}) and (\ref{eq:UDevol}), is just $2(\gamma_{\nu}+\gamma_{\ell})$,  being the pure evolution rate of $Tr(\NN\LL)$ and Det$^{\frac{1}{3}}(\NN\LL)$. Thus two more invariants follow, which we call $\Ili$ and $\Ilj$ respectively, having definitions in terms of $\NN$ and $\LL$ analogous to those in Eqs.~(\ref{eq:Iq1}) and (\ref{eq:Iq2}).

For completeness, we present here other exact one-loop evolution invariants of the SM. The $T$-dependence cancels in the ratio of any corresponding pair of purely-evolving quark and lepton observables, leaving only a dependence on gauge couplings, $g_i$ ($i=1..3$). The one-loop RGEs for the $g_i$ in the SM (at high energies) are \cite{Machacek1}:
\begin{equation}\label{eq:couplings}
\frac{dg_1}{dt}= \frac{41}{6} g_1^3,\quad \frac{dg_2}{dt}= -\frac{19}{6} g_2^3,\quad \frac{dg_3}{dt}= -7 g_3^3.
\end{equation}
Thus eg.~using Eq.~(\ref{eq:UDevol}), 
together with 
its leptonic analogue and Eq.~(\ref{eq:couplings}),
we have that:
\begin{equation}\label{eq:Iq3}
\Iqk\equiv\frac{\Det(\UU\DD)}{\Det(\NN\LL)}g_1^{-\frac{96}{41}}g_3^{-\frac{96}{7}}
\end{equation}
is also an exact one-loop evolution invariant.

We note that by combining Eqs.~(\ref{eq:ourset9}) and
(\ref{eq:ourset10}), to form the pure-evolving
$\Det(\UU\DD)$, we have effectively removed one
independent evolution equation from the complete set,
Eqs.~(11). Thus, we may add the (independent) Jarlskog commutator \cite{Jarlskog1,Jarlskog2} which also has a pure RGE \cite{Athanasiu}:
\begin{equation}\label{eq:jqdet}
\frac{d}{dt} \ln(\Det[\UU,\DD]) = 3[2(\gamma_u+\gamma_d)+ Tr(\UU) + Tr(\DD) ]
\end{equation}
and likewise for the leptons. Noting the definition of $T$, Eq.~(\ref{Eq:T}), and using Eqs.~(\ref{eq:UDevol}), (\ref{eq:jqdet}), their leptonic analogues, and Eq.~(\ref{eq:couplings}), we find another RG invariant:
\begin{equation}\label{eq:I7}
\Ia\equiv \frac{\Det^3[\UU,\DD]\Det[\NN,\LL]}{\Det^{3}(\UU\DD) \Det^{\frac{5}{4}}(\NN\LL)} g_1^{-\frac{81}{82}} g_2^{\frac{81}{38}} .
\end{equation}

Using Eqs.~(\ref{eq:couplings}), two more RG invariants can be constructed from gauge couplings alone:
\begin{align}
\Ib &\equiv \frac{6}{41} g_1^{-2} + \frac{6}{19} g_2^{-2},\label{eq:I8}\\
\Ic &\equiv \frac{6}{41} g_1^{-2} + \frac{1}{7} g_3^{-2}.\label{eq:I9}
\end{align}
Finally, we note the SM RGE of the Higgs vacuum expectation value, $v$ \cite{Eur}:
\begin{equation}\label{eq:vev}
\frac{dv}{dt}=v\left(-T + \frac{3}{4} g_1^2 + \frac{9}{4} g_2^2\right).
\end{equation}
Since its product with any Yukawa coupling gives a mass term, we have that if we use mass matrices directly, rather than Yukawa matrices, the $T$- and $g_2$-dependences of the $\gamma_i$, Eq.~(\ref{Eq:gammas}), are exactly cancelled  leaving only the dependences on $g_1$ and $g_3$. Thus, using $v$ together with purely-evolving quantities, and the gauge couplings, allows the construction of other RG invariants, eg.
\begin{equation}\label{eq:Iqdv}
\Iqv\equiv \Det^{\frac{1}{3}}(\UU\DD)\, v^4 g_1^\frac{4}{41} g_3^{-\frac{32}{7}}.
\end{equation}
Of course, only one of these invariants involving $v$ is independent of the set already defined.

\section{Evaluation}
In constructing our RG invariants, we have used only four of the variables defined in Eq.~(\ref{eq:inv}), namely $\Nd$, $\Ld$, $\NL$ and $\NiLi$.
While ${\rm Det}(\UU\DD)=\Nd\Ld$ is simply the product of all six eigenvalues, variables of the form
$Tr(\UU^n\DD^m)$ depend also on the mixing matrix elements. It is easy to 
show that such 
quantities are simple mass moment transforms \cite{Moriond} of the ``$P$-matrix''  \cite{Pmatrix} of transition probabilities $|V_{\alpha i}|^2$. Writing $u=m_u^2/v^2$, etc., with analogous expressions for the charge $-\frac{1}{3}$ quarks:
\begin{align}
&Tr(\UU^n\DD^m) \notag \\
&\text{\;}=\rowvec {u^n}{c^n}{t^n} \cdot
\mat{|V_{ud}|^2}{|V_{us}|^2}{|V_{ub}|^2}
        {|V_{cd}|^2}{|V_{cs}|^2}{|V_{cb}|^2}
        {|V_{td}|^2}{|V_{ts}|^2}{|V_{tb}|^2}\cdot
\colvec {d^m}{s^m}{b^m} \notag \\
\label{eq:trln}
&\text{\;}=\sum_{\alpha i}m_{\alpha}^{2n} m_{i}^{2m}|V_{\alpha i}|^2/v^{2(m+n)}\quad \forall\, m,n,
\end{align}
(with  $\alpha=u,c,t$ and $i=d,s,b$) in which terms, the flavour-symmetry property is manifest. We may now expand our new RG invariants explicitly.
From Eq.~(\ref{eq:Iq1}):
\begin{align}
\Iqi&=\sum_{\alpha i}\frac{m_{\alpha}^2 m_{i}^2 |V_{\alpha i}|^2}{(m_u m_c m_t m_d m_s m_b)^{\frac{2}{3}}} \notag \\
\label{eq:Iq1expand}
&=\sum_{\alpha\neq\beta\neq\gamma, i\neq j\neq k}\left ( \frac{m_{\alpha}^2}{m_{\beta}m_{\gamma}}\frac{m_{i}^2}{m_j m_k} \right )^{\frac{2}{3}}|V_{\alpha i}|^2.
\end{align}
From Eq.~(\ref{eq:Iq2}):
\begin{align}
\Iqj&=(m_u m_c m_t m_d m_s m_b)^{\frac{2}{3}}\sum_{\alpha i}m_{\alpha}^{-2} m_{i}^{-2}|V_{\alpha i}|^2 \notag \\
\label{eq:Iq2expand}
&=\sum_{\alpha\neq\beta\neq\gamma, i\neq j\neq k}\left ( \frac{m_{\beta}m_{\gamma}}{m_{\alpha}^2}\frac{m_j m_k}{m_{i}^2} \right )^{\frac{2}{3}}|V_{\alpha i}|^2.
\end{align}
Analogous formulae obtain for the leptonic RG invariants, $\Ili$ and $\Ilj$. 

From Eq.~(\ref{eq:Iq3}):
\begin{equation}\label{eq:Iq3expand}
\Iqk= \frac{m_u m_c m_t m_d m_s m_b}{m_1 m_2 m_3 m_e m_{\mu} m_{\tau}} g_1^{-\frac{96}{41}} g_3^{-\frac{96}{7}},
\end{equation}
while from Eq.~(\ref{eq:I7}):
\begin{align}\label{eq:I7expand}
\begin{split}
\Ia=&J_q^3 f^3(u)f^3(d)\times J_{\ell}f(\nu)f(\ell)\\
&\times(y_1 y_2 y_3 y_e y_{\mu}y_{\tau})^{-\frac{1}{4}} g_1^{-\frac{81}{82}} g_2^{\frac{81}{38}},
\end{split}
\end{align}
with $f(u)=(m_t^2-m_c^2)(m_c^2-m_u^2)(m_t^2-m_u^2)/(m_t^2m_c^2m_u^2)$ and
similar definitions for the charge $-\frac{1}{3}$ quarks, and the leptons.  The $y_{\nu}$ and $y_{\ell}$ are the eigenvalues of $\NN$ and $\LL$.

\begin{figure}[htb]
\includegraphics[scale=0.6]{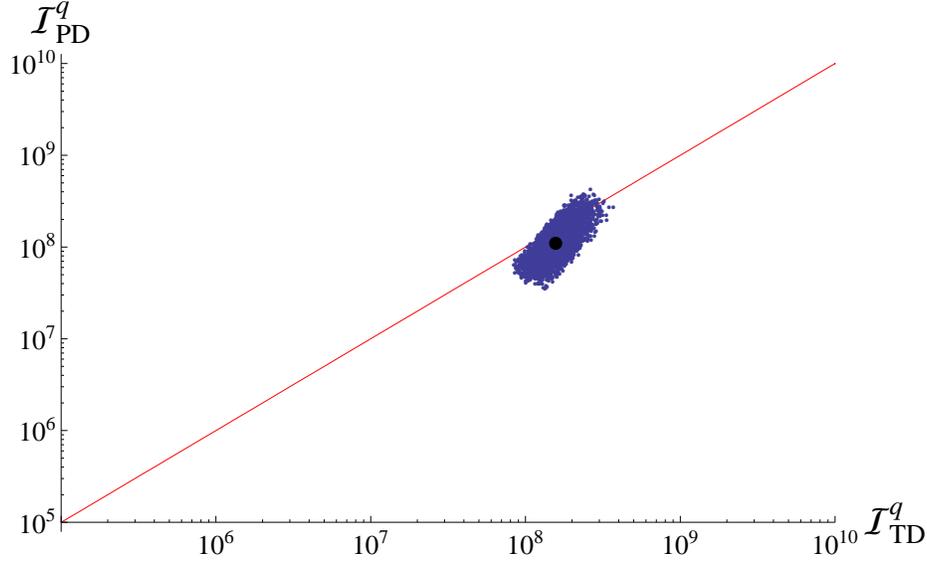}
\caption{\label{fig:inv} The black point shows the values of the RG invariants $\Iqi$ and $\Iqj$ found using quark masses from \cite{XingMass} and measured values of the CKM mixings (all renormalised to $M_Z$). The cluster of points indicates the range allowed by experimental and theoretical uncertainties. 
The straight line shows the hypothesis $\Iqi=\Iqj$ suggested by the data.}
\end{figure}

For brevity, we limit the following discussion to $\Iqi$ and $\Iqj$, the RG invariants constructed only from quark Yukawa matrices. Using the experimental values of the quark masses \cite{XingMass}, and the Wolfenstein  parameters \cite{Wolf}, $\lambda, A, \rho$ and $\eta$ for the CKM matrix, we  find  both invariants to be of the order of $10^8$, as shown in Fig.~\ref{fig:inv}, with their ratio $\Iqptd= 0.7^{+1.1}_{-0.4}$, consistent with unity.
The strongly hierarchical quark masses and the small CKM mixing angles mean that each of them is dominated by a single leading term. We find at  next-to-leading order in small quantities (small mass ratios and $\lambda^2$):
\begin{align}
\label{eq:invapprox1}
&\Iqi\approx \left(\frac{m_t}{m_u}\frac{m_t}{m_c}\frac{m_b}{m_d}\frac{m_b}{m_s}\right)^{\frac{2}{3}}\left(1+\lambda^2 \left(\frac{m_c}{m_t}\frac{m_s}{m_b}\right)^2\right), \\ 
\label{eq:invapprox2}
&\Iqj\approx \left(\frac{m_t}{m_u}\frac{m_c}{m_u}\frac{m_b}{m_d}\frac{m_s}{m_d}\right)^{\frac{2}{3}}(1-\lambda^2).
\end{align}
Since for $\Iqi$, Eq.~(\ref{eq:invapprox1}), the leading term is several orders of magnitude larger than the next-to-leading term, we conclude that the combination $\left(m_t^2 m_b^2/m_u m_c m_d m_s\right)^{\frac{2}{3}}$ is itself invariant to a very good approximation.
At next-to-leading order, the ${\cal O}(1)$ invariant ratio 
is:
\begin{equation}
\frac{\Iqj}{\Iqi}\approx\left(m_c^2 m_s^2/ m_t m_u m_b m_d\right)^{\frac{2}{3}}(1-\lambda^2).
\label{ratioInv}
\end{equation}

It is well-known that from the weak scale to the GUT scale, the various quark masses evolve by typically 55-65\% \cite{XingMass}. The different mass ratios, on the other hand, vary at a slower rate, eg.~$m_b / m_s$ changes by  $\sim$16\% and $m_s / m_d$ by  $\sim$1.8\%. As a check on our analysis, we have numerically solved Eqs.~(\ref{eq:rg1}) and (\ref{eq:rg2}) together 
with the RG equations for the gauge couplings,  Eq.~(\ref{eq:couplings}), and verified that our RG invariants do not evolve at all.
We have similarly verified that the leading terms of our RG invariants given in Eqs.~(\ref{eq:invapprox1}) and (\ref{eq:invapprox2}) change by 0.05\% or less. 

\section{Interpretation}
While the Yukawa coupling matrices  $\UU$ and $\DD$ separately have the mixed and coupled evolutions given by Eqs.~(\ref{eq:hermsqrg1}) and (\ref{eq:hermsqrg2}), 
it is an interesting feature, apparently peculiar to the SM, 
that the eigenvalues, $\lambda_i$, of the {\em product} matrix $\UU\DD$ \footnote{There is no significance to the choice of the product order $\UU\DD$ over $\DD\UU$, since the eigenvalues are the same in each case. All our results are equally applicable to both cases.} have pure evolutions with common rate, leaving the eigenvalue {\em ratios} RG-invariant. This follows since $\NL=Tr(\UU\DD)$ and $\Det(\UU\DD)$ with pure RGEs given in Eqs.~(\ref{eq:ourset5}) and (\ref{eq:UDevol}), 
are simply the order-one and order-three coefficients in the eigenvalue equation of the matrix $\UU\DD$, while $\NiLi=Tr(\UU\DD)^{-1}$, with pure evolution given by  Eq.~(\ref{eq:ourset6}), is simply $P(\UU\DD)/\Det(\UU\DD)$ where:
\begin{equation}
P(\UU\DD)\equiv\frac{1}{2}\left[Tr^2(\UU\DD)-Tr(\UU\DD)^2\right]\label{Eq:Pdef}
\end{equation}
($=\lambda_1\lambda_2+\lambda_2\lambda_3+\lambda_3\lambda_1$)
is the corresponding order-two coefficient. From Eqs.~(\ref{eq:ourset5}), (\ref{eq:ourset6}) and  (\ref{eq:UDevol}), we thus see that each of the coefficients in the eigenvalue equation of $\UU\DD$ has a pure RGE with an evolution rate which is simply given by the order of the coefficient times the same basic rate, $2(\gamma_u+\gamma_d)$. Since the three eigenvalues of  $\UU\DD$ are all order-one in terms of these coefficients via the formula for the roots of a cubic, it follows that they also have 
pure RGEs with common evolution rate  $2(\gamma_u+\gamma_d)$. We thus conclude that the ratios of the eigenvalues of $\UU\DD$, $\lambda_i/\lambda_j$ ($i \neq j$), are also each RG invariants (although clearly they are not individually flavour-symmetric).

In the case of a strong hierarchy, $\Iqi\approx(\lambda_3^2/\lambda_1\lambda_2)^{\frac{1}{3}}$ while $\Iqj\approx(\lambda_2\lambda_3/\lambda_1^2)^{\frac{1}{3}}$. These invariants could, a priori, have taken any values in nature, when, in fact, each is ${\cal O}(10^{8})$. While it is an undoubted mystery why they should be so large, it is also a puzzle why they should be so nearly equal to each other - the proximity to unity of their observed ratio, $\Iqptd\simeq0.7^{+1.1}_{-0.4}$ (see Fig.~\ref{fig:inv}), represents a significant fine-tuning of SM parameters. 
Moreover, if this ratio were exactly unity, then the spectrum of the product matrix $\UU\DD$ would be geometric, ie.
\begin{equation}
\Iqi=\Iqj(={\cal I},~{\rm say}) \Rightarrow\lambda_3/\lambda_2=\lambda_2/\lambda_1\approx
{\cal I},
\label{Eq:geom}
\end{equation}
relations which are then valid at all scales.

It is tempting to postulate that some New Physics requires the spectrum of the matrix $\UU\DD$ to be exactly geometric, $\Iqptd\equiv1$, at some (presumably high) energy scale; the data are consistent with this hypothesis, since, as we have shown, if it is geometric at one scale, then it is geometric at all scales, as long as the one-loop RGEs of the SM remain a good approximation. It has long been known \cite{Davidson} that the separate spectra of the charge $+\frac{2}{3}$ and charge $-\frac{1}{3}$ quarks are approximately geometric: $m_c^2/(m_um_t)\sim{\cal O}(1)$, $m_s^2/(m_dm_b)\sim{\cal O}(1)$. In contrast to our hypothesis, such separate relations are not RG-invariant 
and are therefore a priori less interesting and more difficult to test against data.

We consider briefly why the SM admits RG invariants constructed from only the Yukawa couplings. It can be seen from Eqs.~(\ref{eq:hermsqrg1}) and (\ref{eq:hermsqrg2}),  
that the mixed parts of the evolution equations for the Yukawa coupling matrices $\UU$ and $\DD$ have balanced positive and negative coefficients. These are exploited in the evolution of the product $\UU\DD$ where these terms  cancel  on taking the trace of simple powers. The existence of balanced coefficients in the SM can be traced back to the use of the conjugate Higgs for the Yukawa couplings of the charge $\frac{2}{3}$ quarks, by contrast with the MSSM and the 2HDM, which use independent  Higgs fields in each charge sector, resulting in mixed evolutions \cite{Athanasiu} with coefficients all having the same sign, so that no such cancellation is possible. 

\section{Summary}
We have recast the SM RG equations using flavour-symmetric weak-basis invariant functions of the Yukawa coupling matrices, leading to the identification of exact one-loop RG invariants in the SM. We have identified two such  invariants involving quark Yukawas alone, and two similar ones for leptons in the case of Dirac neutrinos. The SM seems at least somewhat unusual in allowing such RG invariants - we have not been able to find  any in the MSSM or 2HDM. 
Despite the fact that the evolutions of $\UU$ and $\DD$ are coupled and mixed, the weak-basis invariants of their product matrix $\UU\DD$ have pure evolutions with a rate simply proportional to their order, so that its eigenvalue-ratios are RG-invariant, and are furthermore experimentally observed to be consistent with a geometric spectrum.

This work was supported by the UK Science and Technology Facilities Council (STFC). Two of us (PFH and RK) acknowledge the hospitality of the Centre for Fundamental Physics (CfFP) at the Rutherford Appleton Laboratory. We also acknowledge useful comments by Steve King and Roman Zwicky.

\providecommand{\noopsort}[1]{}\providecommand{\singleletter}[1]{#1}

\end{document}